\documentclass[12pt]{iopart}
\usepackage{iopams}
\begin{document}

\title{A path to Balmer's formula: the Pythagorean search for simplicity and harmony}
\author{D V Red\v zi\' c}

\address{Retired from Faculty of Physics, University of Belgrade, PO
Box 44, 11000 Beograd, Serbia} \eads{\mailto{redzic@ff.bg.ac.rs}}

\begin{abstract}
A derivation of Balmer's formula is presented, guided by the
principles of simplicity and harmony.
\end{abstract}
%\pacs{XXX}
%\maketitle

\section{Introduction}
It appears that the crucial point in the development of quantum
physics was a lecture on the spectral lines of hydrogen given by the
mathematician Johann Jakob Balmer on 25 June 1884 before the
Naturforschende Geselschaft in Basel [1]. The wavelengths of the
four visible hydrogen lines, $H_{\alpha}$ (red), $H_{\beta}$
(blue--green), $H_{\gamma}$ (blue) and $H_{\delta}$ (violet), were
precisely measured by {\AA}ngstr\"{o}m [2] to be 6562.1 {\AA},
4860.74 {\AA}, 4340.1 {\AA} and 4101.2 {\AA}, respectively, and it
was a long-standing puzzle for spectroscopists if there was any
regularity behind these numbers. In his lecture Balmer stated, ``So
I gradually arrived at a formula which can apply, at least for these
four lines, as an expression of a law by which their wavelengths are
represented with startling precision.'' The formula is given by the
strikingly simple series,

\begin{equation}
\frac {m^2} {m^2 - 2^2}\times 3645.6 \mbox{{\AA}}\, , \qquad \qquad
m = 3,4,5, ...
\end{equation}
where 3645.6 {\AA} is, in Balmer's terminology, ``fundamental number
of hydrogen'' ({\it Grundzahl}).

The importance of Balmer's formula for the subsequent development of
atomic spectrum theory and quantum physics has been well documented
in literature (e.g., [3]-[9]). However, Balmer does not reveal how
he obtained the formula. While some reconstructions of Balmer's
argument have been attempted [3,5,8,10], it appears that his path to
the celebre formula remains obscure. In this paper, instead of
attempting to reconstruct Balmer's original argument, we derive the
formula in the Balmerian spirit, through the Pythagorean search for
simplicity and harmony.

\section{Derivation of Balmer's formula}
The first insight into the relationships between the wavelengths of
the four main hydrogen lines is due to Stoney [11], who noticed that
ratios of the wavelengths of the first, second and fourth line are
expressed, very accurately, through integers:

\begin{equation}
\frac {H_{\alpha}} {H_{\beta}} = \frac {27}{20}\, , \qquad \frac
{H_{\alpha}} {H_{\delta}} = \frac {8}{5}\, .
\end{equation}
Eqs. (2) obviously imply

\begin{equation}
20 H_{\alpha} = 27 H_{\beta} = 32 H_{\delta} = K \, ,
\end{equation}
where $K$ is a constant for which Stoney gives the value 131277.14
{\AA}.\footnote [1] {Stoney used slightly different wavelengths,
namely, {\AA}ngstr\"{o}m's wavelengths given above corrected for
atmospheric refraction, $H_{\alpha} = 6563.93 \mbox{{\AA}},
H_{\beta} = 4862.11 \mbox{{\AA}},  H_{\delta} = 4102.37
\mbox{{\AA}}$.} Thus

\begin{equation}
H_{\alpha} = \frac {K}{20}\, \quad H_{\beta} = \frac {K}{27}\, ,
\quad  H_{\delta} = \frac {K}{32}\, ,
\end{equation}
i.e. the wavelengths have a common factor $K$, being ``overtones''
of the ``fundamental tone'' $K$. Unfortunately, being unable to
include the remaining visible line, $H_{\gamma}$, into this
regularity, Stoney left this {\it acoustical} line of research.

Balmer [1] reached the same conclusions independently  (or perhaps
he knew of Stoney's work? [3,5,8]), noting that the direct analogy
with acoustics does not yield ``a clearer insight.'' But he stresses
that, ``nevertheless, the idea was obvious that there had to be a
simple formula with whose help the wavelengths of the four indicated
hydrogen lines could be represented.'' Balmer specifies:
``{\AA}ngstr\"{o}m's very precise measurements of the four hydrogen
lines made it possible to seek out a common factor for their
wavelengths that had a most possible simple ratios of numbers to the
wavelengths.'' But, except for the empty sentence, ``So I gradually
arrived at a [valid] formula,'' Balmer offers no explanation of how
he deduced it.\footnote [2] {In Reference [12], where the results of
Reference [1] were republished to get better visibility, Balmer
``omitted all the explanatory material that shows the connection of
his work with the theory of overtones. Instead, his formula appears
to come from nowhere ...'' [8].} Consistent with the Pythagorean
perspective, Balmer's {\it Ansatz} must have been that the
wavelengths equal a common factor multiplied by as simple as
possible irreducible fractions, as suggested by the preserved drafts
of Reference [1] (see [10]). Thus, the ratios of the observed
{\AA}ngstr\"{o}m wavelengths,

\begin{equation}
 \frac {H_{\alpha}} {H_{\beta}}
\approx 1.350 = \frac {27}{20}\, ,
\end{equation}

\begin{equation}
\frac {H_{\alpha}} {H_{\gamma}} \approx 1.512 = \frac {189}{125}\, ,
\end{equation}

\begin{equation}
\frac {H_{\alpha}} {H_{\delta}} \approx 1.600 = \frac {8}{5}\, ,
\end{equation}
should be recast into complex fractions, with irreducible fractions
in their numerators and denominators. With this in mind, we express
the above ratios as

\begin{equation}
 \frac {H_{\alpha}} {H_{\beta}}
 = \frac {27}{20} = \frac {3 \times 9}{5 \times 4} = \frac {3 \times 9}{10 \times 2}\, ,
\end{equation}

\begin{equation}
\frac {H_{\alpha}} {H_{\gamma}} = \frac {189}{125} = \frac {9 \times
21}{5 \times 25} = \frac {27 \times 7}{5 \times 25} = \frac {3
\times 63}{5 \times 25}\, ,
\end{equation}

\begin{equation}
\frac {H_{\alpha}} {H_{\delta}} = \frac {8}{5} = \frac {9 \times
8}{9 \times 5} = \frac {3 \times 8}{3 \times 5}\, .
\end{equation}

Inspection of eqs. (8) - (10) reveals that only two irreducible
fractions can correspond to $H_{\alpha}$ and comply with simplicity
requirement, 9/5 and 3/5. Obviously, the first choice yields

\begin{equation}
H_{\alpha} = B9/5\, , \quad H_{\beta} = B4/3\, , \quad H_{\gamma} =
B25/21\, , \quad H_{\delta} = B9/8\, ,
\end{equation}
and the second choice yields

\begin{equation}
H_{\alpha} = B^*3/5\, , \quad H_{\beta} = B^*4/9\, , \quad
H_{\gamma} = B^*25/63\, , \quad H_{\delta} = B^*3/8\, ,
\end{equation}
where $B$ and $B^*$ are the respective common factors.\footnote [3]
{Choosing, for example, 3/2 for $H_{\alpha}$ yields 10/9, 125/126
and 15/16  for $H_{\beta}$, $H_{\gamma}$ and $H_{\delta}$,
respectively, which appears to be less simple than the choices (11)
and (12). It can be easily verified that all other choices for
$H_{\alpha}$ (3/4, 9/4, 3/10, 9/10, etc), lead to sets of cumbersome
irreducible fractions. }

Thus, search for simplicity singled out two eligible sets of
irreducible fractions for the four hydrogen lines. Now, following
the Pythagoreans (Balmer included), we attempt to find the right set
searching for harmony.

This feat was accomplished by Balmer, who noticed that in eqs. (11),
the first and the third irreducible fraction can be written in the
form

\begin{equation}
\frac {9}{5} = \frac {3^2}{3^2 - 2^2}\, , \qquad \frac {25}{21} =
\frac {5^2}{5^2 - 2^2}\, ,
\end{equation}
which suggested that the second and the fourth fraction in (11) were
nothing but

\begin{equation}
\frac {4}{3} = \frac {4^2}{4^2 - 2^2}\, , \qquad \frac {9}{8} =
\frac {6^2}{6^2 - 2^2}\, .
\end{equation}
The value of the constant factor $B$ is obtained as the arithmetic
mean of $H_{\alpha}5/9$, $H_{\beta}3/4$, $H_{\gamma}21/25$ and
$H_{\delta}8/9$, which gives $B = 3645.6$ {\AA}.\footnote [4]
{Balmer notes, ``on various grounds it seems to me likely'' that a
more general formula applies, $Bm^2/(m^2 - n^2)$, where $m$ and $n$
are integers. But, as is well known, Balmer is wrong here, the
correct formula reads $(B/4) m^2n^2/(m^2 - n^2)$.}

On the other hand, no harmony could be recognized in the set
expressed by eqs. (12).

Thus we arrived at Balmer's formula guided by the principles of
simplicity and harmony.

\section{Concluding comments}
It appears that Balmer felt that his formula provides a peep through
the keyhole into the invisible. As a devoted Pythagorean, always in
quest for simplicity, harmony and symmetry, he probably believed
that the hydrogen spectrum is a picture of the harmonious structure
of the hydrogen atom. In Balmer's prophetic words, ``hydrogen, whose
atomic weight is by far the smallest among the atomic weights of all
substances known to date and characterizes it as the simplest
chemical element, [...], seems more destined than any other body to
open up new avenues of research into the nature of matter and its
properties.'' [1]

Closing this tribute to Balmer, we cannot help recalling Galileo's
sentence from {\it Discorsi} that nature herself in carrying out all
her works employs ``the primary means, the simplest, the easiest''
({\it mediis primis, simplicissimis, facillimis}).

\end{document}